\documentclass[final]{l4dc2026}

\title[Learning Recommendations under Imperfect Human Compliance with RLDP]{Learning Recommendations under Imperfect Human Compliance with Record-Level Differential Privacy}
\usepackage{times}
\usepackage{titlesec}
\usepackage{verbatim}
\usepackage{amsmath,amssymb,amsfonts}
\usepackage{mathtools}
\usepackage{caption}
\usepackage{natbib}
\usepackage{enumitem}
\usepackage{titlesec}

\titlespacing*{\section}{0pt}{1.1ex plus 0.2ex minus 0.2ex}{0.8ex plus 0ex}
\titlespacing*{\subsection}{0pt}{0.8ex plus 0.0ex minus 0.2ex}{0.6ex plus 0ex}

\usepackage{etoolbox}
\BeforeBeginEnvironment{proof}{\vspace{-4pt}}
\AfterEndEnvironment{proof}{\vspace{-3pt}}

\BeforeBeginEnvironment{theorem}{\vspace{-4pt}}
\AfterEndEnvironment{theorem}{\vspace{-4pt}}
\BeforeBeginEnvironment{lemma}{\vspace{-4pt}}
\AfterEndEnvironment{lemma}{\vspace{-4pt}}

    \author{%
 \Name{Yuwen Ma} \Email{yuwen.ma.24@ucl.ac.uk}\\
 \addr University College London
 \AND
 \Name{Nicola Bastianello} \Email{n.bastianello@ucl.ac.uk}\\
  \addr University College London
  \AND
 \Name{Sarah Spurgeon} \Email{s.spurgeon@ucl.ac.uk}\\
 \addr University College London%
  \AND
  \Name{Boli Chen} \Email{boli.chen@ucl.ac.uk}\\
 \addr University College London\vspace{-1em}
}

\begin{document}

\maketitle
\vspace{-0.5em}
    \begin{abstract}%

This paper investigates human-in-the-loop recommendation systems, where effectiveness relies on human compliance and privacy concerns restrict the use of historical response data. The proposed framework models the implemented action as a random response to the recommendation and formulates the recommendation mechanism as a decision-dependent optimisation problem under an unknown compliance distribution. A batched Gaussian process upper confidence bound algorithm is then adapted to learn from locally perturbed historical actions, providing privacy for individual raw records. The analysis establishes record-level differential privacy and derives high-probability simple regret bounds when the compliance distribution is unknown. These bounds characterise how privacy noise, batch size, and behavioural information affect the guaranteed quality of the learned recommendation. In particular, a known upper bound on the non-compliance rate enables the recommendation space to be screened before any private data are
queried. Simulations of a fixed-state car-following problem with synthetic historical driver responses demonstrate the effectiveness of the proposed algorithm and verify the theoretical findings.
\end{abstract}


\section{Introduction}
\begin{figure}[!htbp]
    \centering
    \vspace{-1.5em}
    \includegraphics[width=0.5\linewidth]{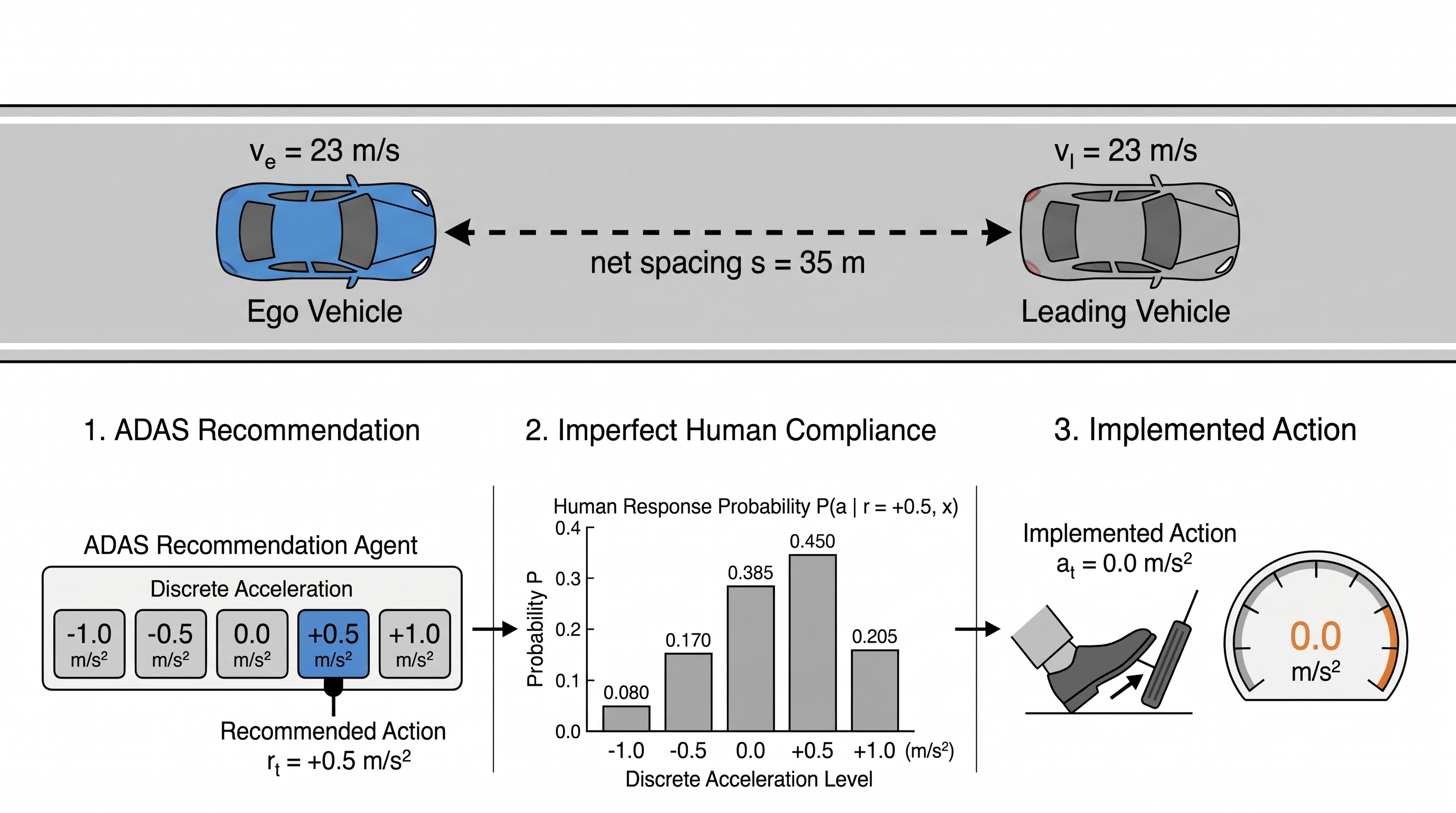}\\[-2ex]
    \captionsetup{font=small}
    \caption{Illustration of imperfect human compliance in a car-following scenario.}
    \label{fig0}
\end{figure}

Decision-making in cyber–physical–human systems (CPHS) has attracted growing interest in control and optimisation \citep{annaswamy2023cyber}, spanning applications from assisted driving \citep{sentouh2018driver} to demand response \citep{cavraro2026demand}. In many human-centric CPHS architectures, a recommendation agent provides operational advice, while humans determine the actions ultimately applied to the physical system. The recommendation may specify a target acceleration in an advanced driver assistance system (ADAS), a thermostat setpoint in residential demand response, or an irrigation volume in agriculture. Since the human does not always follow advice exactly (see Figure~\ref{fig0} for an ADAS car-following example), the effectiveness of the recommendation agent critically depends on the human response it induces.

To capture this dependence, let \(r\) denote the action recommended by the recommendation agent, \(a\) the action implemented by the human, and \(x\in\mathbb R^m\) the context of the physical system. Here, $r,a\in\mathcal{R}$, and \(\mathcal R\subset\mathbb{R}^n\) denotes a finite, discrete space. Imperfect human compliance is modelled by a conditional probability distribution $a\sim P(\cdot\mid r,x)$,
which describes how the implemented action depends on $r$ and $x$. Given a fixed system context \(x\) and a system cost \(\ell(a,x): \mathcal{R}\rightarrow\mathbb{R}\), the recommendation problem can be formulated as
\begin{equation}\label{e1}
    \min_{r\in\mathcal R} J(r),
\qquad
J(r)=\mathbb E_{a\sim P(\cdot\mid r,x)}
\bigl[\ell(a,x)\bigr].
\end{equation}The recommendation performance is thus influenced by the distribution $P(\cdot\mid r,x)$.

However, in practice, the human compliance distribution \(P(\cdot\mid r,x)\) is generally unknown to the recommendation agent. For a fixed operating context \(x\), a trusted cloud $\mathcal C_h$ stores historical actions grouped by the recommendations under which they were recorded. The recommendation agent queries these logs sequentially to learn the expected system loss \(J(r)\). To protect individual records, each sampled action is perturbed by the cloud using a differentially private mechanism before transmission \citep{dwork2014algorithmic}. The privatised queries therefore reflect both variability in human responses and randomness introduced by the privacy mechanism.

Under imperfect compliance, the expected loss of a recommendation depends on how often it is followed and which actions are taken when it is ignored. \citet{grand2024best} describe deviations through a baseline policy, while \citet{bang2026route} assume uniform choice among alternative routes. But, when alternative actions incur different losses, their individual probabilities matter. This motivates modelling non-compliance via an unknown compliance distribution, analogous to the treatment response formulation in instrument-armed bandits \citep{kallus2018instrument}. In CPHS, \citet{dave2024framework} learn the human response model directly from observed actions. However, access to such actions raises privacy issues. Related work has investigated learning from privatised user feedback in interactive recommendation systems \citep{zheng2020locally}, as well as privacy-preserving data aggregation and state inference in human-in-the-loop systems \citep{ cortes2016differential}.  Motivated by these studies, this paper formulates a general recommendation learning problem: \textit{how can a recommendation agent learn an optimal recommendation from privacy-protected historical data when the compliance distribution is unknown?}

To solve this problem, a batched Gaussian process upper confidence bound (GP-UCB) algorithm \citep{srinivas2009gaussian} is adapted to learn a recommendation from perturbed historical actions using debiased loss estimates. The main contributions are summarised as follows.
\begin{enumerate}[nosep]
    \item Departing from prior work that treats compliance and privacy in isolation, a human-in-the-loop recommendation framework that jointly accounts for imperfect human compliance and record-level differential privacy (RLDP) is proposed. The dependence of human actions on recommendations is captured by a conditional compliance distribution, leading to a decision-dependent stochastic optimisation problem at the recommendation agent.
    
    \item The analysis establishes a RLDP guarantee for the full sequence of adaptive queries and derives high-probability simple regret bounds. These results show how human behaviour and privacy noise enter the learning regret. In particular, a known upper bound on the non-compliance rate yields a screening rule that eliminates provably suboptimal recommendations before any private records are queried.

    \item A fixed-state car-following case study evaluates the proposed algorithm using synthetic historical driver responses and a one-step driving loss. Across different compliance distributions, the algorithm achieves low simple regret and outperforms a compliance-unaware recommendation. The simulations also illustrate the effects of privacy noise and compliance on simple regret and optimal-recommendation identification.
\end{enumerate}

\section{Problem Setup and Preliminaries}

This section presents the system setup and assumptions for problem~\eqref{e1}, followed by preliminaries on Gaussian process regression and the local differential privacy (LDP) mechanism.

\subsection{System Setup and Assumptions}\label{se2.1}
Following \eqref{e1}, $\mathcal{R}\! =\! \{\alpha_1, \ldots, \alpha_K\}$ denotes a finite action space. The cloud maintains historical datasets $\mathcal D=\bigcup_{i=1}^{K}\mathcal D_i,\mathcal{D}_i\!=\!\{(a_{i,n},\alpha_i,x)\}_{n=1}^{N_i}$, where $a_{i,n}\in\mathcal R$ is the $n$-th recorded action implemented on the physical system in response to $\alpha_i$.

For each query from the recommendation agent, the cloud returns a locally privatised response sampled from $\mathcal{D}$ to protect any individual raw record within the historical dataset. To help the recommendation agent mitigate the effects of privacy noise, the cloud also provides the noise distribution, including its mean and variance, without revealing individual noise realisations. The recommendation agent knows the loss function $\ell$, the space $\mathcal R$, and the fixed context $x$, but not the compliance distributions $\{P(\cdot|\alpha_i,x)\}_{i=1}^{K}$. The following assumption is required. 

\textbf{Assumption 1} The compliance distributions $\{P(\cdot|\alpha_i,x)\}_{i=1}^{K}$ remain unchanged throughout the collection of the historical dataset $\mathcal{D}$. Each query for recommendation $\alpha_i$ is answered by sampling uniformly with replacement from $\mathcal D_i$, using fresh sampling randomness at each query.

Under this setup, this work aims to learn an optimal recommendation for \eqref{e1} from locally privatised query responses and derive regret bounds, while limiting what the recommendation agent or an external adversary can infer about any individual raw record within the historical dataset.

\subsection{Gaussian Process Regression and Confidence Bounds}\label{se2.2}

Gaussian process regression (GPR) is a nonparametric method for learning an unknown function from noisy observations, providing a posterior mean and associated uncertainty across the domain. GP-UCB then combines these quantities to construct an upper confidence bound that guides the sequential selection of recommendations by balancing exploration and exploitation. The mathematical formulation is as follows.

Let \(k:\mathcal R\times\mathcal R\to\mathbb R\) be a positive semidefinite kernel satisfying \(k(r,r)\leq 1\),  \(\forall r\in\mathcal R\), and \(\mathcal H_k\) denotes its associated reproducing kernel Hilbert space (RKHS). Given a prior mean function $\mu_0: \mathcal{R} \to \mathbb{R}$, consider an unknown target function $f \in \mathcal{H}_k$ satisfying $\Vert{}f - \mu_0\Vert{}_{\mathcal{H}_k} \le B$. At each time iteration \(t\), a recommendation \(r_t\in\mathcal R\) is selected based on the prior history \(\mathcal F_{t-1}\), and an observation $
y_t = f(r_t) + \omega_t$ is revealed, where the noise \(\omega_t\) is conditionally \(\nu\)-sub-Gaussian, meaning \(\mathbb E[\mathrm{e}^{u\omega_t}\mid\mathcal F_{t-1}] \leq \mathrm{e}^{u^{2}\nu^{2}/2}\), \(\forall u\in\mathbb R\).

Given observations \(\mathcal O_t=\{(r_i,y_i)\}_{i=1}^{t}\), let \(\mathbf y_{1:t}\!=\![y_1,\ldots,y_t]^\top\), \(\boldsymbol{\mu}_{0,1:t}=[\mu_0(r_1),\ldots,\mu_0(r_t)]^\top\), \(\mathbf k_t(r)\!\!=\![k(r_1,r),\ldots,k(r_t,r)]^\top\), and $K_{t}=[k(r_i,r_j)]_{i,j= 1}^t$ be the kernel matrix. With a GP prior mean $\mu_0(r)$ and unit regularisation, the posterior mean and variance at any \(r\in\mathcal R\) are given by
\begin{equation}\label{e2}
    \mu_t(r) \!=\! \mu_0(r)\!+\!\mathbf k_t(r)^\top(K_t\!+\!\mathbf I_t)^{-1}(\mathbf y_{1:t}\!-\!\boldsymbol{\mu}_{0,1:t}), \quad 
s_t^2(r)\! = \! k(r,r)\! - \!\mathbf k_t(r)^\top(K_t\!+\!\mathbf I_t)^{-1}\mathbf k_t(r),
\end{equation}
with \(s_0^2(r)=k(r,r)\), and \(\mathbf I_t\) is the identity matrix. Following the GP-UCB framework \citep{srinivas2009gaussian}, a high-probability uniform confidence bound is derived from \eqref{e2} by adapting self-normalised concentration inequalities to the RKHS setting~\citep{abbasi2011improved}. Specifically, for any \(\delta_{\mathrm c}\in(0,1)\), with probability at least \(1-\delta_{\mathrm c}\),
\begin{equation}\label{e3}
    |f(r)-\mu_{t-1}(r)| \leq b_t\,s_{t-1}(r), \qquad \forall r\in\mathcal R,\; \forall t\geq 1,
\end{equation}
where the confidence multiplier is given by
$
b_t = B + \nu\sqrt{2\left(\mathcal I_{t-1} + \log\frac{1}{\delta_{\mathrm c}}\right)},
$ and \(\mathcal I_t = \frac12\log\det(\mathbf I_t+K_t)\) denotes the information gain after \(t\) evaluations (\(\mathcal I_0=0\)).

\subsection{Local Differential Privacy Mechanism}\label{se2.3}

Let $\mathcal{S}_{i,t} = \{(\bar{a}_{i,t,j}, \alpha_i, x)\}_{j=1}^{M}$ denote a batch drawn independently and uniformly with replacement from $\mathcal{D}_i$ at iteration $t$. To protect individual records in $\mathcal{D}$, each sampled action is locally perturbed before transmission to the recommendation agent for GPR updates and GP-UCB selection.

\textbf{Definition 1 \citep{dwork2014algorithmic}} A randomised mechanism \(\mathcal M:\mathcal U\to\mathcal V\) satisfies \((\varepsilon,\delta)\)-DP if, for all \(u,u'\in\mathcal U\) and all measurable subsets \(\mathcal S\subseteq\mathcal V\), $$
\Pr\!\left(\mathcal M(u)\in\mathcal S\right)
\leq
e^\varepsilon
\Pr\!\left(\mathcal M(u')\in\mathcal S\right)+\delta,$$
where \(\varepsilon>0\) and \(\delta\in[0,1)\) denote the privacy parameters. When \(\delta=0\), $\mathcal{M}$ satisfies pure \(\varepsilon\)-DP.


\begin{lemma}[\citealp{le2013differentially}]\label{le1} Let $\epsilon >0$, $\delta\in (0,\frac{1}{2})$ and $\mathcal{Q}(x)=\frac{1}{\sqrt{2\pi}}\int_x^\infty\mathrm{e}^{-\frac{u^2}{2}}du$. Consider a function \(q:\mathcal U\to\mathbb R^n\) with \(\ell_2\)-sensitivity \(\Delta_2(q) = \sup_{u,u'\in\mathcal U}\|q(u)-q(u')\|_2\). 
    The mechanism \(\mathcal M(u) = q(u) + \eta\), $\eta\sim\mathcal{N}(0,\sigma^2)$, satisfies \((\varepsilon,\delta)\)-DP, where $\sigma \ge \frac{\Delta_{2}(q)}{\sqrt{{\mathcal{Q}^{-2}(\delta)}+2\varepsilon}-\mathcal{Q}^{-1}(\delta)}$. 
\end{lemma}

\section{Compliance-Aware Recommendation Algorithm with Privacy Preservation}\label{se3}

This section presents the proposed algorithm, summarised in Algorithm~\ref{alg1}. Under Assumption~1, it applies batched GP-UCB to solve~\eqref{e1} via privatised queries, using a privacy channel to debias loss estimates and updating a GP model for recommendation selection. The main steps follow.

\textbf{Local Sampling and Private Queries.} At iteration $t$, let $r_t = \alpha_i$ ($i \in \{1, \dots, K\}$) be the recommendation selected by the recommendation agent. The cloud draws a batch $\mathcal{S}_{i,t}$ from $\mathcal{D}_i$ and transmits the perturbed batch $\tilde{\mathcal{S}}_{i,t} = \{(\tilde{a}_{i,t,j}, \alpha_i, x)\}_{j=1}^M$ to the recommendation agent, where
\begin{equation}
    \tilde{a}_{i,t,j} = {a}_{i,t,j} + \eta_{i,t,j}, \quad \eta_{i,t,j} \overset{\text{i.i.d.}}{\sim} \mathcal{N}(0, \sigma^2 I_n), \quad j = 1, \dots, M, \label{e4}
\end{equation}
with noise samples independent of the sampled actions. Upon receiving $\tilde{\mathcal{S}}_{i,t}$, the recommendation agent projects each perturbed action onto $\mathcal{R}$ via the nearest-neighbour operator:
\begin{equation}
    z_{t,j} = \Pi_{\mathcal{R}}(\tilde{a}_{i,t,j}) \triangleq \alpha_{k^*(\tilde{a}_{i,t,j})}, \quad k^*(u) \triangleq \min \Big\{ \arg\min_{k \in \{1,\dots,K\}} \|u - \alpha_k\|_2 \Big\}. \label{e5}
\end{equation}

\textbf{Privacy Channel and Loss Debiasing.} The perturbation in \eqref{e4} and projection in \eqref{e5} induce a recommendation-independent privacy channel with column-stochastic transition matrix $W_\sigma \in \mathbb{R}^{K \times K}$ ($\mathbf{1}_K^\top W_\sigma = \mathbf{1}_K^\top$), whose entries are
\begin{equation}
    [W_\sigma]_{kl} = \Pr(\Pi_{\mathcal{R}}(\alpha_l + \eta) = \alpha_k) = \int_{\mathcal{V}_k} \frac{1}{(2\pi\sigma^2)^{n/2}} e^{-\frac{\|u - \alpha_l\|_2^2}{2\sigma^2}} \,\mathrm{d}u,  \quad \eta\sim\mathcal N(\mathbf 0,\sigma^2\mathbf I_n), \label{e6}
\end{equation}
where $\mathcal{V}_k = \{u \in \mathbb{R}^n : \Pi_{\mathcal{R}}(u) = \alpha_k\}$ denotes the Voronoi projection region associated with $\alpha_k$. Let $\boldsymbol{\ell} = [\ell(\alpha_1, x), \dots, \ell(\alpha_K, x)]^\top$. Assuming $W_\sigma$ is invertible, the recommendation agent precomputes the debiased loss vector $\tilde{\boldsymbol{\ell}}_\sigma = [\tilde{\ell}_{\sigma,1}, \dots, \tilde{\ell}_{\sigma,K}]^\top$ prior to the online iterations by solving
\begin{equation}
    W_\sigma^\top \tilde{\boldsymbol{\ell}}_\sigma = \boldsymbol{\ell}. \label{e7}
\end{equation}
By mapping each projected sample $z_{t,j} = \alpha_k$ to $\tilde{\ell}_\sigma(z_{t,j}, x) = [\tilde{\ell}_\sigma]_k$, the debiased batch loss at iteration $t$ is computed as
\begin{equation}
    \hat{J}_t^{\mathrm{deb}} = \frac{1}{M} \sum_{j=1}^M \tilde{\ell}_\sigma(z_{t,j}, x). \label{e8}
\end{equation}

\RestyleAlgo{boxruled}
\setlength{\algomargin}{1.5em}
\begin{algorithm2e}[H]
\caption{Batched GP-UCB with debiased private feedback}
\label{alg1}

\LinesNumbered
\SetAlgoVlined

\KwIn{
Action set $\mathcal R$, context $x$,
losses $\{\ell(\alpha_i,x)\}_{i=1}^K$,
noise level $\sigma$, batch size $M$, horizon $T$,
kernel parameters $h,\rho$, confidence parameter $\delta_c$, prior pair $(\mu_0, B)$
}
\KwOut{Recommendation $\hat r_T$}

Compute $L_{\min}$, $L_{\max}$, $D$, and $c$\;

Construct $W_\sigma$ by \eqref{e6} and solve
\eqref{e7}\;

Compute $\tilde D_\sigma$ and $\nu_\sigma$ by
\eqref{e10};

Set $s_0(r)=1$ for all $r\in\mathcal R$,
and set $\mathcal I_0=0$\ (default prior: $\mu_0(r) = 0, B = \frac{\sqrt{K}}{2\sqrt{\rho}}$);

\For{$t=1,\ldots,T$}{
    Compute $b_t$ and select $r_t$
    by \eqref{e12}\;

    Record
    $q_t=\mu_{t-1}(r_t)-b_t s_{t-1}(r_t)$\;

    For $r_t=\alpha_i$, the cloud samples $M$ actions
    independently with replacement from $\mathcal D_i$
    and transmits their perturbed values by \eqref{e4}\;

    Project the received values by \eqref{e5} and
    calculate $y_t$ by \eqref{e8} and
    \eqref{e9}\;

    Add $(r_t,y_t)$ to the GP data and update
    $\mu_t$, $s_t$, and $\mathcal I_t$\;
}

Set $\tau\in\arg\max_{1\leq t\leq T}q_t$ and
$\hat r_T=r_\tau$\;

\KwRet $\hat r_T$\;
\end{algorithm2e}

\textbf{GP-Based Recommendation Selection and Update.} Define $L_{\min} \triangleq \min_i \ell(\alpha_i, x)$, $L_{\max} \triangleq \max_i \ell(\alpha_i, x)$, $D \triangleq L_{\max} - L_{\min}$, and $c \triangleq (L_{\min} + L_{\max})/2$. If $D=0$, every recommendation is optimal, so this case is omitted here. For the non-trivial case $D > 0$, minimising $J(r)$ is equivalent to maximising the normalised objective $f(r)$, with the corresponding GP observation $y_t$ given by
\begin{equation}
    f(r) = \frac{c - J(r)}{D}, \qquad y_t = \frac{c - \hat{J}_t^{\mathrm{deb}}}{D}. \label{e9}
\end{equation}
To calibrate the GP confidence bound under debiased batch observations, set the span of the debiased loss vector and the effective noise scale (corresponding to $\nu$ in Section \ref{se2.2}) as
\begin{equation}
    \tilde{D}_\sigma = \max_k [\tilde{\ell}_\sigma]_k - \min_k [\tilde{\ell}_\sigma]_k, \qquad \nu_\sigma = \frac{\tilde{D}_\sigma}{2D\sqrt{M}}. \label{e10}
\end{equation}
Adopt the regularised Gaussian kernel
\begin{equation}
    k(r, r') = (1 - \rho) e^{-\frac{\|r - r'\|_2^2}{2h^2}} + \rho \mathbf{1}_{\{r = r'\}}, \quad h > 0, \; 0 < \rho \le 1. \label{e11}
\end{equation}
The GP is initialised with $s_0(r)=1$ for all $r\in\mathcal R$ and a prior pair $(\mu_0,B)$ satisfying $\|f-\mu_0\|_{\mathcal H_k}\le B$. Without prior compliance information, set $\mu_0(r)=0$ and $B=\frac{\sqrt K}{2\sqrt\rho}$. At each iteration $t$, using the confidence multiplier $b_t$ from Section \ref{se2.2} (with $\nu = \nu_\sigma$ and confidence parameter $\delta_c \in (0,1)$), the algorithm selects the recommendation $r_t$ and records its pessimistic score $q_t$:
\begin{equation}
    r_t \in \arg\max_{r \in \mathcal{R}} \big[\mu_{t-1}(r) + b_t s_{t-1}(r)\big], \qquad q_t = \mu_{t-1}(r_t) - b_t s_{t-1}(r_t). \label{e12}
\end{equation}
After updating the posterior mean \(\mu_t(r)\) and standard deviation \(s_t(r)\) with $y_t$ via \eqref{e2}, the algorithm returns $\hat{r}_T = r_\tau$ after $T$ iterations, where $\tau \in \arg\max_{1 \le t \le T} q_t$.

\section{Privacy and Regret Analysis}

This section establishes privacy and regret guarantees for
Algorithm~\ref{alg1}. The GP confidence bound \eqref{e3} requires a
bounded RKHS norm and conditionally sub-Gaussian observation noise.
These conditions are first examined for the empirical objective induced by the dataset $\mathcal{D}$.

Let $p(\alpha_i)=[P(\alpha_1\mid\alpha_i,x),\ldots,
P(\alpha_K\mid\alpha_i,x)]^\top$ denote the compliance distribution
under $\alpha_i$, and let $P=[p(\alpha_1),\ldots,p(\alpha_K)]$ be
the compliance transition matrix. For the fixed dataset $\mathcal D$,
let $\hat p_i$ be the empirical action distribution in $\mathcal D_i$
and define $J_{\mathcal D}(\alpha_i)=\boldsymbol{\ell}^\top\hat p_i$.
For $r_t=\alpha_i$, Assumption~1 implies that $\{z_{t,j}\}_{j=1}^M$ are conditionally i.i.d.\ given
$\mathcal F_{t-1}$, with distribution $W_\sigma\hat p_i$. By
\eqref{e7},
\[
\mathbb E[\tilde{\ell}_\sigma(z_{t,j},x)\mid\mathcal F_{t-1}]
=\tilde{\boldsymbol{\ell}}_\sigma^\top W_\sigma\hat p_i
=\boldsymbol{\ell}^\top\hat p_i
=J_{\mathcal D}(r_t).
\]
Thus, $\hat J_t^{\mathrm{deb}}$ is conditionally unbiased for
$J_{\mathcal D}(r_t)$. For $D>0$, define
$f_{\mathcal D}(r)=(c-J_{\mathcal D}(r))/D$. By \eqref{e9},
$\mathbb{E}[y_t-f_{\mathcal D}(r_t)]=0$. Since the
debiased losses are conditionally independent and each lies in an
interval of length $\tilde D_\sigma$, Hoeffding's Lemma
\citep{hoeffding1963probability} gives the conditional sub-Gaussian
parameter $\nu_\sigma=\tilde D_\sigma/(2D\sqrt M)$ in \eqref{e10}.
If $\hat p_i=p(\alpha_i)$ for every $i$, then $J_{\mathcal D}=J$
and $f_{\mathcal D}=f$.

For the default prior $\mu_0=0$, let $K_{\mathcal R}$ be the kernel
matrix evaluated on $\mathcal R$ and let
$\boldsymbol f_{\mathcal D}=[f_{\mathcal D}(\alpha_1),\ldots,
f_{\mathcal D}(\alpha_K)]^\top$. The diagonal regulariser in
\eqref{e11} gives $\lambda_{\min}(K_{\mathcal R})\ge\rho$.
Since $|f_{\mathcal D}(r)|\le 1/2$ for all $r\in\mathcal R$, $
\|f_{\mathcal D}-\mu_0\|_{\mathcal H_k}
=\sqrt{\boldsymbol f_{\mathcal D}^{\top}
K_{\mathcal R}^{-1}\boldsymbol f_{\mathcal D}}
\le \frac{\sqrt K}{2\sqrt\rho}=B.
$

Consequently, the confidence bound \eqref{e3} holds uniformly throughout Algorithm~\ref{alg1} with probability at least $1-\delta_c$. The privacy and regret guarantees are established in the following subsections.

\subsection{Record-Level Privacy Guarantee of Algorithm~\ref{alg1}}
Let $\Delta_{\mathcal{R}} \triangleq \max_{a, a' \in \mathcal{R}} \|a - a'\|_2$ denote the diameter of the action space. The RLDP guarantee is established as follows.
\begin{theorem}\label{th3}
Under Assumption~1, select $\delta_0 \in (0,\frac{1}{2})$ and $\sigma > 0$ in \eqref{e4} such that $\varepsilon_0 \triangleq \frac{\Delta_{\mathcal{R}}^2}{2\sigma^2} + \frac{\Delta_{\mathcal{R}}}{\sigma} \mathcal{Q}^{-1}(\delta_0)$. For each $i \in \{1, \dots, K\}$, every raw action record $a_{i,n}$ in the historical dataset $\mathcal{D}_i = \{(a_{i,n}, \alpha_i, x)\}_{n=1}^{N_i}$ satisfies at least $(\varepsilon_{i,T}, \delta_{i,T})$-RLDP over $T$ iterations of Algorithm~\ref{alg1}, where
\begin{equation}\label{e13}
    \varepsilon_{i,T} = M T \log\!\left(1 + \frac{e^{\varepsilon_0}-1}{N_i}\right),
    \qquad
    \delta_{i,T} = \frac{M T \delta_0}{N_i}.
\end{equation}
\end{theorem}

\begin{proof}
Consider a single sample draw at iteration $t \le T$ when $r_t = \alpha_i$. Define the perturbation mechanism in \eqref{e4} as $\mathcal{M}(a) \triangleq a + \eta$ for any $a \in \mathcal{R}$. Since the $\ell_2$-sensitivity of releasing an action in $\mathcal{R}$ is $\Delta_{\mathcal{R}}$, Lemma~\ref{le1} ensures that $\mathcal{M}$ satisfies $(\varepsilon_0, \delta_0)$-DP with respect to each sampled action $\bar{a}_{i,t,j}$. 

Next, analyse the privacy guarantee for any specific raw record $a_{i,n} \in \mathcal{D}_i$ under a single uniform draw with replacement. Define the subsampled mechanism $\mathcal{M}_{\mathrm{s}}(\mathcal{D}_i) \triangleq \mathcal{M}(\bar{a}_{i,t,j})$, which draws an action $\bar{a}_{i,t,j}$ uniformly at random from $\mathcal{D}_i$ and then applies the base mechanism $\mathcal{M}$. Let $\mathcal{D}_i$ and $\mathcal{D}_i'$ be two neighbouring datasets of size $N_i$ differing only in the $n$-th record ($a_{i,n}$ versus $a_{i,n}'$). By Assumption~1, a single draw selects the $n$-th record with probability $\frac{1}{N_i}$, and selects from the remaining $N_i - 1$ identical records with probability $1 - \frac{1}{N_i}$ (inducing a common output distribution $Q$ under both $\mathcal{D}_i$ and $\mathcal{D}_i'$). Thus, for any measurable set $\mathcal{S}$, the output of $\mathcal{M}_{\mathrm{s}}$ satisfies
\begingroup
\small
\begin{align*}
    \Pr(\mathcal{M}_{\mathrm{s}}(\mathcal{D}_i) \!\in\! \mathcal{S}) 
   \! = \!\frac{N_i - 1}{N_i} Q(\mathcal{S})\! +\! \frac{1}{N_i} \Pr(\mathcal{M}(a_{i,n})\! \in\! \mathcal{S})
    \!\le\! \frac{N_i - 1}{N_i} Q(\mathcal{S})\! +\! \frac{1}{N_i} \Big(e^{\varepsilon_0} \Pr(\mathcal{M}(a_{i,n}')\! \in \!\mathcal{S})\! +\! \delta_0\Big),
\end{align*}
\endgroup
where the inequality follows from the $(\varepsilon_0, \delta_0)$-DP property of $\mathcal{M}$. Applying the standard privacy amplification bound by subsampling \citep{balle2018privacy} yields $\Pr(\mathcal{M}_{\mathrm{s}}(\mathcal{D}_i) \in \mathcal{S}) \le e^{\varepsilon_{{i}}} \Pr(\mathcal{M}_{\mathrm{s}}(\mathcal{D}_i') \in \mathcal{S}) + \delta_{i}$, confirming that $\mathcal{M}_{\mathrm{s}}$ satisfies $(\varepsilon_i, \delta_i)$-RLDP with $\varepsilon_i = \log\!\big(1 + \frac{e^{\varepsilon_0}-1}{N_i}\big)$ and $\delta_i = \frac{\delta_0}{N_i}$.

Finally, since each query to $\mathcal{D}_i$ invokes $\mathcal{M}_{\mathrm{s}}$ independently $M$ times and $\mathcal{D}_i$ is queried at most $T$ times, $\mathcal{M}_{\mathrm{s}}$ is executed no more than $M T$ times on $\mathcal{D}_i$. By the sequential composition theorem \citep{dwork2014algorithmic} and the post-processing immunity of the projection operator \eqref{e5}, the overall privacy loss over $T$ iterations is bounded by $(\varepsilon_{i,T}, \delta_{i,T})$ in \eqref{e13}.
\end{proof}

While the bound in~\eqref{e13} depends on $T$, it serves as an \textit{a priori} worst-case guarantee by assuming $\mathcal{D}_i$ is queried in every iteration. In practice, because query decisions are adaptive, the actual number of queries made to~$\mathcal{D}_i$ by termination is a trajectory-dependent quantity that may be strictly smaller, yielding a tighter \textit{ex-post} realised privacy loss $(\varepsilon'_{i,T},\delta'_{i,T})$. 

\subsection{Regret Bounds and Behavioural Insights for Algorithm~\ref{alg1}}\label{se4.2}

Let $r^\star\in\arg\min_{r\in\mathcal R}J(r)$. The simple regret of Algorithm~\ref{alg1} after $T$ iterations is defined as $S_T\triangleq J(\hat r_T)-J(r^\star)$. Moreover, let $\gamma_T\triangleq\max_{r_{1:T}\in\mathcal R^T}\mathcal I_T$ denote the maximum information gain over all $T$-step recommendation sequences, and $\kappa(r)\triangleq\Pr(a\neq r\mid r,x)$ the non-compliance rate.

\begin{lemma}[\citealp{srinivas2009gaussian,abbasi2011improved}]\label{le3}For any sequence $\{r_1, \dots, r_T\} \in \mathcal{R}^T$ and horizon $T \ge 1$ under $k(r,r) \le 1$ and $\vert{}\mathcal{R}\vert{} = K$, the cumulative posterior variance and maximum information gain satisfy
$\sum_{t=1}^T s_{t-1}^2(r_t) \le \frac{2\mathcal{I}_T}{\log 2} \le \frac{2\gamma_T}{\log 2}$, and $\gamma_T \le \frac{K}{2}\log\!\left(1 + \frac{T}{K}\right).$
\end{lemma}
\vspace{-0.5em}
\begin{theorem}\label{th5}
Suppose $\hat p_i=p(\alpha_i), \forall i\in\mathcal{R}$ and Assumption 1 hold, $D>0$, and $W_\sigma$ in \eqref{e6} is invertible. Run Algorithm~\ref{alg1} with the GP prior mean $\mu_0$ and parameter $B$ specified by one of the following settings:

\begin{enumerate}[nosep]
    \item \textbf{Unknown compliance distribution.} Set default prior: $\mu_0(r) = 0, B = \frac{\sqrt{K}}{2\sqrt{\rho}}$. \label{th5.1}
    
    \item \textbf{Known non-compliance bound.} Suppose $\kappa(r)\leq\bar\kappa, \forall r\in\mathcal{R}$, where $\bar\kappa\in[0,1]$ is known. Set \label{th5.2}
    \vspace{-0.5em}
\begin{align*}
    g(r)=\frac{c-\ell(r,x)}{D},\quad
\mu_0(r)=(1-\bar\kappa)g(r),\quad
B=\frac{\bar\kappa\sqrt K}{2\sqrt\rho},
\end{align*}
\vspace{-0.5em}
\end{enumerate}
\vspace{-0.5em}

For either setting and any $\delta_c\in(0,1)$, the simple regret of Algorithm~\ref{alg1} satisfies the following bound simultaneously for all $T\geq 1$, with probability at least $1-\delta_c$:
\begin{equation}
\label{e14}
J(\hat r_T)-J(r^\star) \leq 2Db_T\sqrt{\frac{2\gamma_T}{T\log 2}},
\end{equation}
where $
b_T
=
B+\frac{\tilde D_\sigma}{2D\sqrt M}
\sqrt{2\left(\mathcal I_{T-1}+\log\frac{1}{\delta_c}\right)}$, $\mathcal I_{T-1}\leq\gamma_T
\leq\frac K2\log\left(1+\frac TK\right).$
\end{theorem}

\begin{proof}
For the first setting, the preceding analysis establishes $\|f-\mu_0\|_{\mathcal H_k}=\|f\|_{\mathcal H_k}\leq \frac{\sqrt{K}}{2\sqrt{\rho}}$ with $\hat p_i=p(\alpha_i)$. For the second, expanding $f(r)=\sum_{a\in\mathcal R}P(a\mid r,x)g(a)$ yields $f(r)-\mu_0(r)=(\bar\kappa-\kappa(r))g(r)+\sum_{a\neq r}P(a\mid r,x)g(a)$. Since the coefficients are non-negative and sum to $\bar\kappa$, and $|g(a)|\leq 1/2$, $\forall a\in\mathcal R$, it follows that $|f(r)-\mu_0(r)|\leq\bar\kappa/2$. Combined with $K_{\mathcal R}\!\succeq\!\rho I_K$ implies $\|f-\mu_0\|_{\mathcal H_k}\!\leq\!\bar\kappa\sqrt K/(2\sqrt\rho)$. Hence, $\|f-\mu_0\|{\mathcal H_k}\leq B$ holds for either setting with its $B$.

 With $\hat p_i=p(\alpha_i)$, the preceding analysis shows that
$y_t-f(r_t)$ is conditionally $\nu_\sigma$-sub-Gaussian. Applying the confidence bound \eqref{e3} to $f-\mu_0$ ensures that, with probability at least $1-\delta_c$, $|f(r)-\mu_{t-1}(r)|\leq b_t s_{t-1}(r)$ for all $r\in\mathcal R$ and $t\geq 1$. Conditioned on this event, the selection rule \eqref{e12} implies $f(r^\star)\leq \mu_{t-1}(r_t)+b_t s_{t-1}(r_t)=q_t+2b_t s_{t-1}(r_t)$ and $q_t\leq f(r_t)$. Since $\hat r_T=r_\tau$ with $\tau\in\arg\max_{1\leq t\leq T}q_t$ and $b_t\leq b_T$, applying the Cauchy--Schwarz inequality and Lemma~\ref{le3} gives $f(r^\star)-f(\hat r_T)\leq \min_{1\leq t\leq T}(f(r^\star)-q_t)\leq \frac{2b_T}{T}\sum_{t=1}^T s_{t-1}(r_t)\leq 2b_T\sqrt{2\gamma_T/(T\log 2)}$. Multiplying both sides by $D$ establishes \eqref{e14}.
\end{proof}

 Combining Theorem~\ref{th3} and Theorem~\ref{th5} yields some key insights into the interplay between privacy protection, human compliance, and learning performance:
\begin{enumerate}
\item \textbf{Privacy–Suboptimality Trade-off under Imperfect Compliance.} Although exact loss debiasing yields unbiased batch estimates, their concentration depends on the debiased loss span \(\tilde D_\sigma\), which enters the regret bounds in \eqref{e14}. For an invertible \(W_\sigma\), this span satisfies $
\max_{i\ne j}
\frac{|\ell_i-\ell_j|}
{2\Phi\!\left(\|\alpha_i-\alpha_j\|_2/(2\sigma)\right)-1}
\le \tilde D_\sigma
\le
\frac{\sqrt{2}\,\|\ell-c\mathbf 1\|_2}
{s_{\min}(W_\sigma)},$ where \(s_{\min}(W_\sigma)\) is the smallest singular value of \(W_\sigma\). The lower bound is at least$
\sqrt{2\pi}\,\sigma
\max_{i\ne j}
\frac{|\ell_i-\ell_j|}{\|\alpha_i-\alpha_j\|_2}.
$ Thus, the lower bound on \(\tilde D_\sigma\) grows at least linearly with \(\sigma\). The upper bound depends on the conditioning of the privacy channel and may be large when \(W_\sigma\) is ill-conditioned. These bounds characterise the dependence of the stated regret guarantees on privacy noise. The effect of \(\sigma\) on the regret bound also depends on the information gain $\mathcal{I}_t$ of the adaptively selected recommendations, which can vary with the human compliance distribution.

\item \textbf{Guidelines for Batch Size Selection:} Increasing the batch size $M$ shrinks $S_T$ at rate $M^{-1/2}$. However, since $\varepsilon_{i,T}$ scales linearly with $M$ (modulated by the dataset size $N_i$), a larger $M$ accelerates RLDP budget consumption. Let $M_{\mathrm{req}}$ denote the minimum batch size required to guarantee $S_T \le \tau$ (Theorem \ref{th5}), and let $M_{\mathrm{priv}}$ denote the maximum batch size permitted under the $T$-round RLDP budget $(\bar{\varepsilon}, \bar{\delta})$ and dataset size $N_i$ (Theorem \ref{th3}). Then the condition $M_{\mathrm{req}} \le M_{\mathrm{priv}}$ certifies that target accuracy and privacy are jointly attainable. Furthermore, under a fixed total sample budget $MT$, a dominant structural complexity $B$ favours more rounds with smaller batches to prioritise exploration, whereas a large debiased span $\tilde{D}_\sigma$ favours fewer rounds with larger batches to suppress per-step estimation variance.

\item \textbf{Recommendation space screening via $\bar{\kappa}$:} Suppose $\bar{\kappa}$ is known \textit{a priori} to the recommendation agent. For $r_0 \in \arg\min_{r \in \mathcal{R}} \ell(r,x)$ and $\Delta(r) = \ell(r,x) - \ell(r_0,x)$, the bound $J(r) - J(r_0) \ge (1-\bar{\kappa})\Delta(r) - \bar{\kappa} D$ ensures that every minimiser of $J$ resides in the screened set $\mathcal{A} = \{r \in \mathcal{R} : (1-\bar{\kappa})\Delta(r) \le \bar{\kappa} D\}$, computable offline from $\ell$ and $\bar{\kappa}$. In particular, if $(1-\bar{\kappa})\min_{r \ne r_0}\Delta(r) > \bar{\kappa} D$, then $\mathcal{A} = \{r_0\}$, allowing $r_0$ to be returned directly without querying any logs. Otherwise, restricting Algorithm~\ref{alg1} to $\mathcal{A}$ tightens the RKHS norm bound to $B_{\mathcal{A}} = \bar{\kappa}\sqrt{|\mathcal{A}|}/(2\sqrt{\rho})$ and the maximum information gain to $\gamma_T(\mathcal{A}) \le \tfrac{1}{2}|\mathcal{A}|\log(1 + T/|\mathcal{A}|)$.
\end{enumerate}

\begin{remark}
Exact loss debiasing in Theorem~\ref{th5} requires
$\boldsymbol{\ell}\in\operatorname{Range}(W_\sigma^\top)$,
which holds automatically when $W_\sigma$ is invertible.
Under this condition, even if $W_\sigma$ is rank-deficient,
a valid debiased loss vector is
$\tilde{\boldsymbol{\ell}}_\sigma
=(W_\sigma^\top)^\dagger\boldsymbol{\ell}$ (the Moore--Penrose pseudoinverse). Since $W_\sigma^\top\tilde{\boldsymbol{\ell}}_\sigma
=\boldsymbol{\ell}$, the resulting loss estimates remain exactly unbiased. The conclusion of Theorem~\ref{th5} and the regret
bounds retain their stated form, with the debiased loss span
and the corresponding noise parameter computed from $\tilde{\boldsymbol{\ell}}_\sigma
=(W_\sigma^\top)^\dagger\boldsymbol{\ell}$.
\end{remark}
\vspace{-0.5em}

\section{A Case Study in Advanced Driver Assistance Systems}

This section applies the proposed framework to longitudinal driving
advisory, with the ADAS as the recommendation agent, and examines the effects of privacy noise and driver compliance.

\subsection{Mathematical Modelling}
\begingroup
\setlength{\abovedisplayskip}{3pt}
\setlength{\abovedisplayshortskip}{3pt}
\setlength{\belowdisplayskip}{3pt}
\setlength{\belowdisplayshortskip}{3pt}
Consider a highway car-following scenario (see Figure \ref{fig0}) in which an ADAS seeks an
optimal steady-state acceleration recommendation for a representative,
discretised traffic regime (e.g.\ a clustered cruising band of speeds
and inter-vehicle spacings). The context for this regime is fixed as
$
x=(v_e,v_l,s,\theta_d),
$\endgroup
where $v_e$, $v_l$, and $s$ denote the ego speed, leading vehicle speed,
and net spacing, respectively. The parameter vector
$\theta_d=(v_d,s_0,t_h,a_{\mathrm{pref}},b_{\mathrm{pref}},
\delta_{\mathrm{drv}})$ specifies the reference speed, standstill
spacing, time headway, acceleration and braking magnitudes, and model
exponent. Both recommended and executed accelerations belong to the
finite set $\mathcal R=\{\alpha_1,\ldots,\alpha_K\}\subset\mathbb R$,
where $\alpha_1<\cdots<\alpha_K$.

Under the proposed framework, imperfect driver compliance is modelled
by the unknown compliance transition matrix $P$, and
historical responses are collected as described in Section~\ref{se2.1}.

Following the cost structure in \citet[Eq.~(14)]{lot2025eco}, the
driving loss is evaluated at the predicted one-step state:
\begin{equation}
\begin{aligned}
\ell(a,x)=
\delta_{\mathrm{drv}}^2
\left(\frac{v_a}{v_d}-1\right)^2
+\left(\frac{[a]_+}{a_{\mathrm{pref}}}\right)^2
+\left(\frac{[-a]_+}{b_{\mathrm{pref}}}\right)^2+\gamma(v_a)
\frac{(s_a/s_{d,a}-1)^2}
     {(s_a/s_{d,a})^2+1},
\end{aligned}
\label{e15}
\end{equation}
where $[z]_+=\max\{z,0\}$, $s_{d,a}=s_0+t_hv_a$, and
$\gamma(v)=8[(v/v_d)^{\delta_{\mathrm{drv}}}-1]^2$. Assuming constant
leading vehicle speed and ego acceleration over $\Delta t$, the
predicted speed and spacing are $v_a=v_e+a\Delta t$ and
$s_a=s+\Delta t(v_l-v_e)-\frac12a(\Delta t)^2$. Further modelling details are given in \citet{lot2025eco}. The recommendation objective is the expected loss $J(r)=\sum_{k=1}^{K}
\ell(\alpha_k,x)P(\alpha_k\mid r,x)$ as in \eqref{e1}. The optimal recommendation therefore depends on the driver's response.

For the scalar action set, the privacy channel in \eqref{e6} can be
computed using the standard normal cumulative distribution function
$\Phi$. Let $d_i=(\alpha_i+\alpha_{i+1})/2$ for
$i=1,\ldots,K-1$, with $d_0=-\infty$ and $d_K=+\infty$. For
$\sigma>0$, the channel entries are computed as
\begin{equation}
[W_\sigma]_{kl}
=
\Phi\left(\frac{d_k-\alpha_l}{\sigma}\right)
-
\Phi\left(\frac{d_{k-1}-\alpha_l}{\sigma}\right),
\qquad k,l=1,\ldots,K.
\label{e16}
\end{equation}
The ADAS evaluates the loss vector
$\boldsymbol{\ell}=[\ell(\alpha_1,x),\ldots,\ell(\alpha_K,x)]^\top$
and obtains the debiased loss vector by solving
$W_\sigma^\top\tilde{\boldsymbol{\ell}}_\sigma
=\boldsymbol{\ell}$. Algorithm~\ref{alg1} then learns a recommendation using perturbed
actions sampled from the historical dataset, without knowledge of $P$.

\subsection{Simulation Setup and Results}
\begingroup
\setlength{\abovedisplayskip}{5pt}
\setlength{\abovedisplayshortskip}{5pt}
\setlength{\belowdisplayskip}{5pt}
\setlength{\belowdisplayshortskip}{5pt}
The simulations consider the fixed traffic regime
$
v_e=v_l=23~\mathrm{m/s},\, s=35~\mathrm{m},
$
with $v_d=30~\mathrm{m/s}$, $s_0=2~\mathrm{m}$,
$t_h=1.3~\mathrm{s}$, $a_{\mathrm{pref}}=2~\mathrm{m/s^2}$,
$b_{\mathrm{pref}}=3~\mathrm{m/s^2}$,
$\delta_{\mathrm{drv}}=4$, and $\Delta t=1~\mathrm{s}$.
The candidate acceleration set is $\mathcal R=\{-1,-0.5,0,0.5,1\}~\mathrm{m/s^2}$. Then equation~\eqref{e15} gives the corresponding loss vector
$\boldsymbol{\ell}=
[1.2931,\;1.0547,\;0.8858,\;0.8203,\;0.8921]^\top.$

To examine the effect of compliance, the following transition matrices are considered:
\begingroup
\setlength{\arraycolsep}{3pt}
\renewcommand{\arraystretch}{0.9}
\setlength{\abovedisplayskip}{2pt}
\setlength{\abovedisplayshortskip}{2pt}
\setlength{\belowdisplayskip}{2pt}
\setlength{\belowdisplayshortskip}{2pt}
\[
P_{\mathrm{base}}=\begin{bmatrix}
0.450&0.385&0.055&0.385&0.385\\
0.055&0.450&0.055&0.055&0.055\\
0.055&0.055&0.450&0.055&0.055\\
0.055&0.055&0.055&0.450&0.055\\
0.385&0.055&0.385&0.055&0.450
\end{bmatrix}, \ P(\psi)=\psi\mathbf I+\frac{1-\psi}{0.55}(P_{\mathrm{base}}-0.45\mathbf I),
\]
\endgroup
where $\psi\in \Psi=\{0.45,0.65,0.80,0.90\}$ is the compliance probability and $P(0.45)=P_{\mathrm{base}}$. For each $\psi$ and $\alpha_i$, a synthetic historical dataset $\mathcal{D}_i$ of $N_i = 3{,}200{,}000$ records is constructed with action frequencies matching the $i$-th column of the associated matrix $P(\psi)$. Algorithm~\ref{alg1} is evaluated with $T=100$, $M=60$, $h=0.55$, $\rho=0.25$, $\delta_c=0.05$, and candidate Gaussian noise standard deviations $\sigma \in \Sigma=\{0.60, 0.65, 0.70, 0.75, 0.80\}$. Taking $\delta_0 = 0.1$, \eqref{e13} yields $\max_i \delta'_{i,T}\leq 1.875 \times 10^{-4}.$ Each
simulation configuration is averaged over 100 independent Monte Carlo runs. Figures \ref{fig1}, \ref{fig2} summarise the numerical results.

Figure~\ref{fig1}(a) compares the mean terminal regret of Algorithm~\ref{alg1} at $\sigma=0.60$ with the compliance-unaware baseline $r_0=\arg\min_{r\in\mathcal R}\ell(r,x)=0.5\,\mathrm{m/s^2}$ across all $\psi\in \Psi$. Since $r_0$ is computed directly from $\ell(r,x)$ without accessing historical logs, it is unaffected by noise. For Algorithm~\ref{alg1}, the corresponding realised privacy losses $\max_i\varepsilon'_{i,T}$ are $15.5926$, $14.2066$, $13.1671$, and $14.5531$, respectively. These confirm the effectiveness of Algorithm~\ref{alg1} under RLDP requirements. Figure~\ref{fig1}(b) illustrates the impact of the privacy noise $\sigma$ on recommendation performance and privacy protection under $\psi = 0.45$. Shaded bands represent the pointwise $95\%$ confidence intervals across $100$ Monte Carlo runs. Across $\Sigma$, the corresponding realised privacy losses $\max_i \varepsilon'_{i,T}$ are $15.5926$, $4.8344$, $1.8576$, $0.8396$, and $0.4511$, respectively. Notably, increasing $\sigma$ beyond $0.70$ yields diminishing marginal reductions in $\max_i \varepsilon'_{i,T}$ at the expense of a sharp rise in $S_T$. \textit{This reflects the worsening conditioning of $W_\sigma$ at high noise levels, which inflates the debiased loss span $\tilde{D}_\sigma$ and suggests the existence of a $\sigma$ for balancing privacy protection and learning performance.}

Figure~\ref{fig2}(a) shows terminal regret and optimal recommendation identification rates across $\psi\in \Psi$ and $\sigma\in\Sigma$, with $95\%$ Student-$t$ and Wilson intervals, respectively. Under moderate noise, Algorithm~\ref{alg1} achieves effective privacy protection alongside high identification accuracy and low $S_T$. Interestingly, a higher compliance rate $\psi$ may make the optimal recommendation harder to identify, yet simultaneously reduce the cost of misidentification. This occurs because certain compliance distributions flatten $J(r)$ near the optimum. For instance, at $\psi=0.80$, $J(r)$ evaluates to $1.215, 1.077, 0.897, 0.894$, and $0.950$ for $r \in \{-1, -0.5, 0, 0.5, 1\}$, respectively, yielding a marginal optimality gap between $J(0)$ and $J(0.5)$. This narrow gap makes exact best-recommendation identification difficult from finite privatised samples, thereby reducing identification accuracy. Figure~\ref{fig2}(b) compares the simple regret $S_t$ at $\sigma=0.80$ with and without a known non-compliance bound $\bar{\kappa}$. Without prior knowledge of $\bar{\kappa}$, Algorithm~\ref{alg1} searches the full action set $\mathcal{R}$. Given $\bar{\kappa}=1-\psi$, the screening rule in Section~\ref{se4.2} reduces the candidate set to $\mathcal{A}=\{0.5\}$, $\{0,0.5,1\}$, and $\{-0.5,0,0.5,1\}\,\mathrm{m/s^2}$ for $\psi=0.90$, $0.80$, and $0.65$, respectively (For $\psi=0.45$, $\mathcal{A}=\mathcal{R}$). Restricting Algorithm~\ref{alg1} to the screened set $\mathcal{A}$ eliminates redundant exploration and consequently improves learning efficiency.

\begin{figure}[!htbp]
    \centering
    \begin{minipage}[t]{0.34\linewidth}
        \centering
        \vspace{0pt}
        \includegraphics[width=\linewidth]{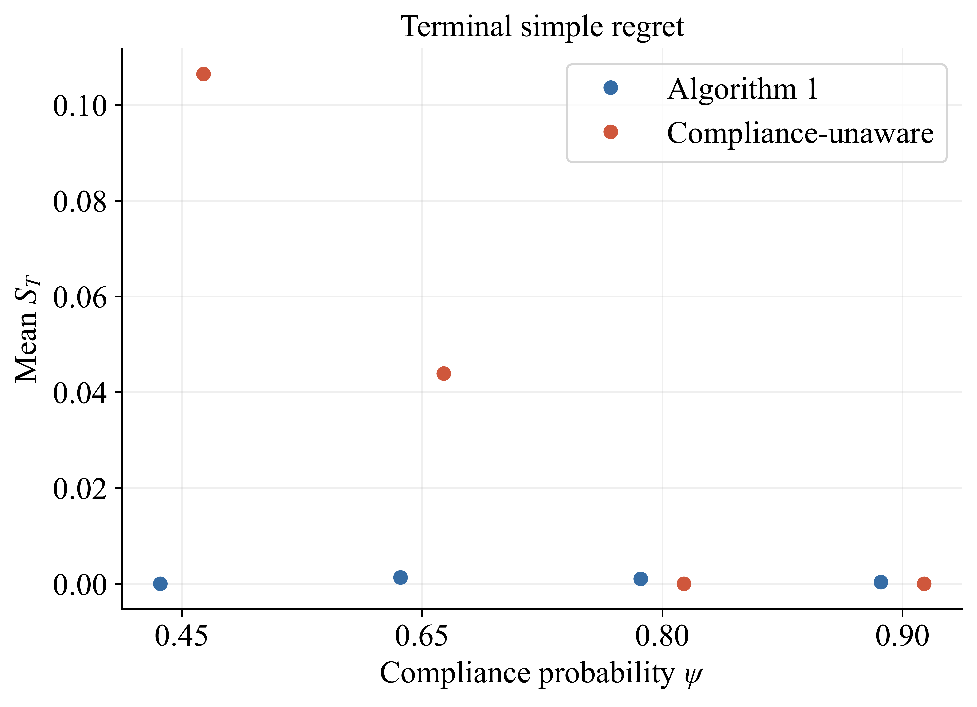}\par
        \smallskip\footnotesize (a) Compliance-aware vs.\ unaware
    \end{minipage}\hfill
    \begin{minipage}[t]{0.66\linewidth}
        \centering
        \vspace{0pt}
        \includegraphics[width=\linewidth]{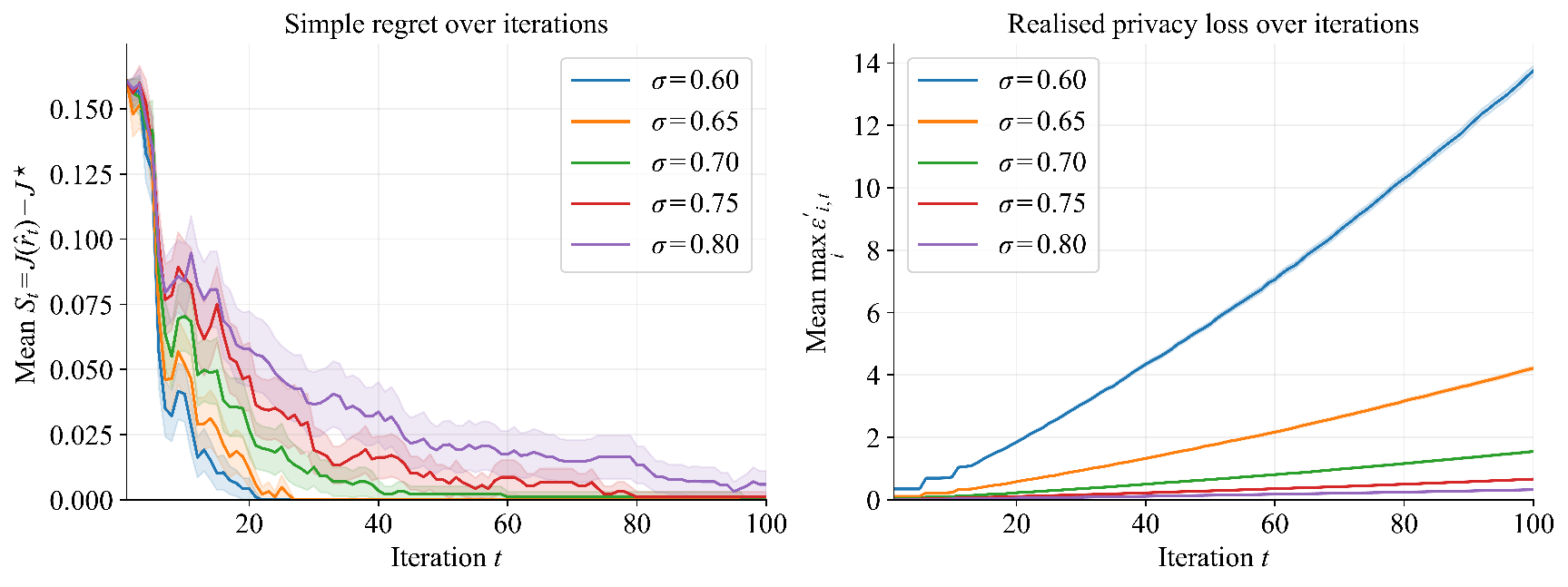}\par
        \smallskip\footnotesize (b) Effect of privacy noise
    \end{minipage}\hfill
    \captionsetup{font=small}
\caption{Effectiveness of Algorithm \ref{alg1} under imperfect human compliance and privacy noise.}
    \label{fig1}
    \vspace{-15pt}
\end{figure}

\begin{figure}[!htbp]
    \centering
    \begin{minipage}[t]{0.604\linewidth}
        \centering
        \vspace{0pt}
        \includegraphics[width=\linewidth]{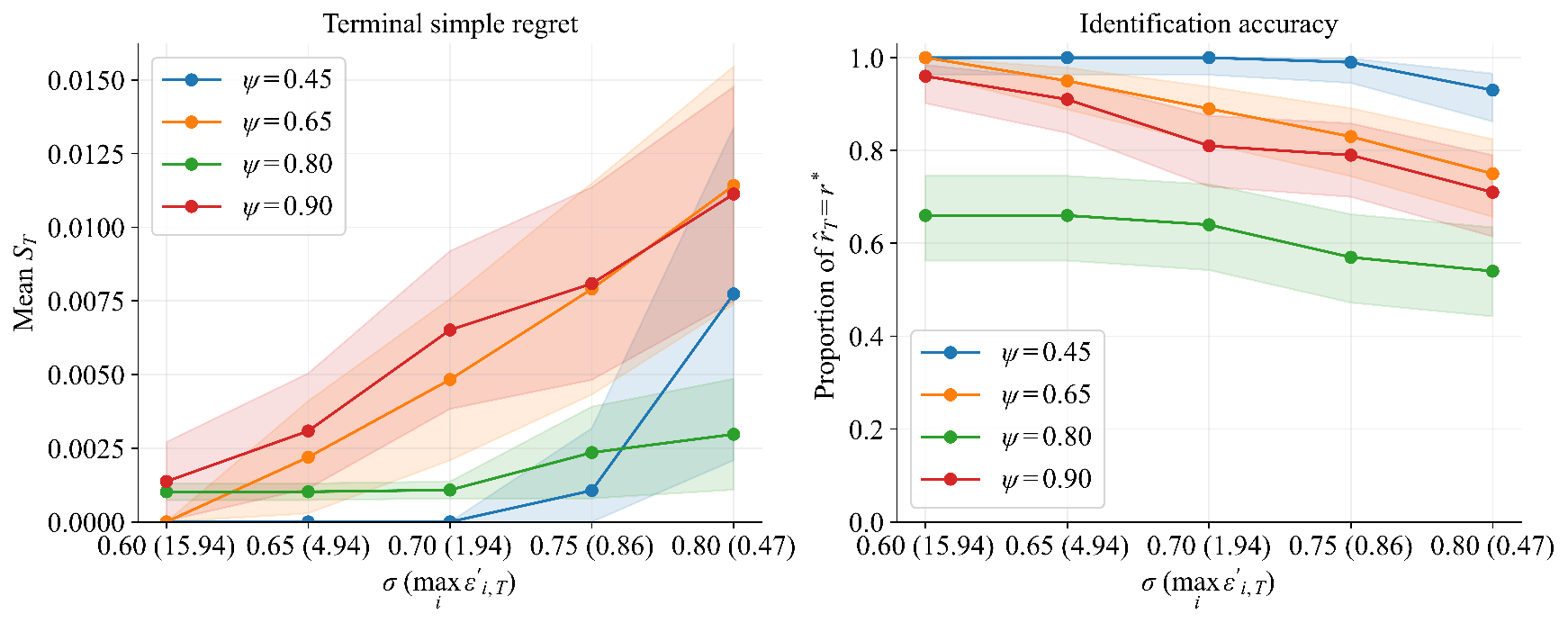} \\[-1ex] \par
        \smallskip\footnotesize (a) Interplay between compliance sharpness and privacy noise
    \end{minipage}\hfill
    \begin{minipage}[t]{0.396\linewidth}
        \centering
        \vspace{0pt}
        \includegraphics[width=\linewidth]{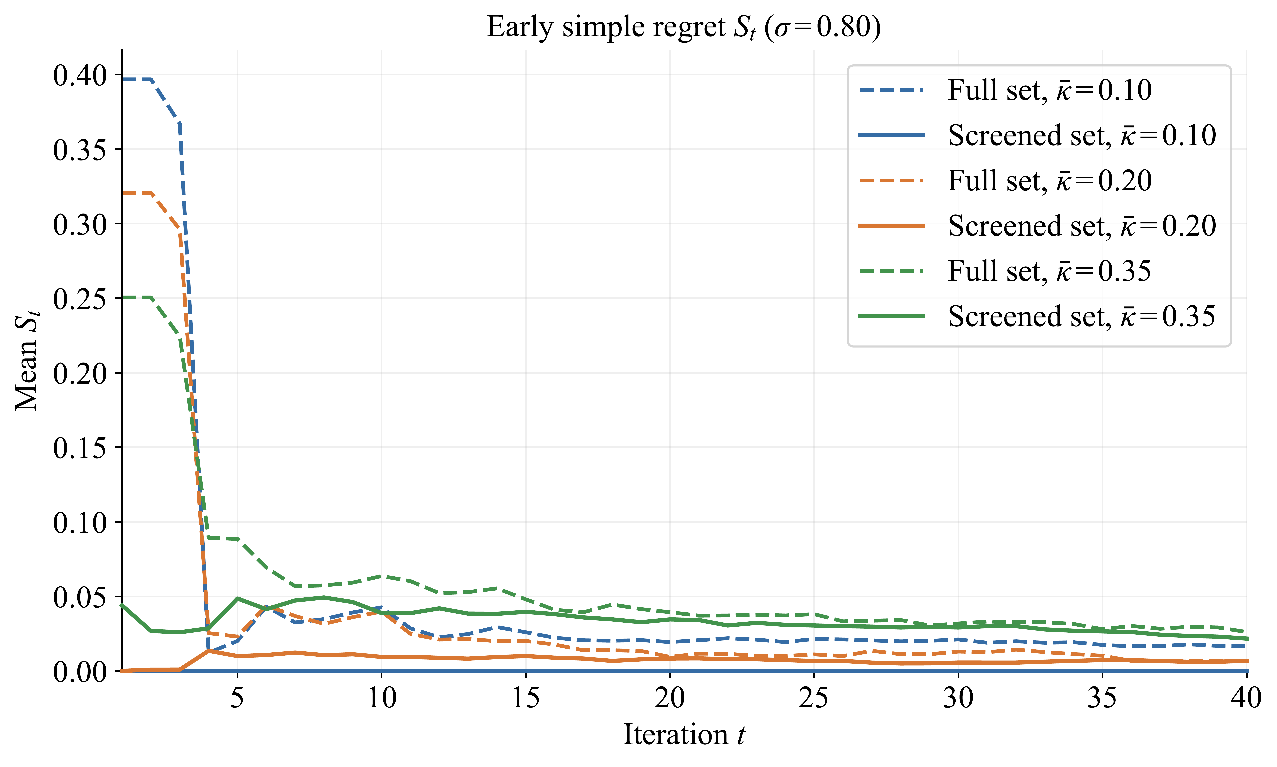}\\[-1ex] \par
        \smallskip\footnotesize (b) Known vs.\ unknown $\bar{\kappa}$
    \end{minipage}  \hfill
    \captionsetup{font=small}
    \caption{Joint effects of compliance-privacy interplay and prior action-space screening.}
    \label{fig2}
     \vspace{-15pt}
\end{figure}

\section{Conclusion}
This paper developed a human-in-the-loop recommendation learning framework under imperfect human compliance. The analysis established privacy and simple regret guarantees and showed how prior compliance information can support screening before private data access. In a car-following simulation, the algorithm achieved low regret across the tested compliance distributions. The framework provides a starting point for studying more general recommendation frameworks with complex human behaviour, evolving physical dynamics and safety constraints.



\bibliography{l4dc2026-sample}

\end{document}